\documentclass[10pt]{ismd08}
\usepackage{graphicx}
\usepackage{graphics}
\usepackage{cite,./mcite}

\newcommand{\lsim}{\,{\buildrel < \over {_\sim}}\,}
\newcommand{\gsim}{\,{\buildrel > \over {_\sim}}\,}
\newcommand {\pt} {\mbox{$p_{\rm t}$}}
\newcommand {\tev} {\mbox{${\rm TeV}$}}
\newcommand {\gev} {\mbox{${\rm GeV}$}}

\newcommand{\dNdeta}{\mbox{${\rm d}N_{\rm ch}/{\rm d}\eta$}}

\newcommand{\ccbar}{\mbox{$\rm c\overline c$}}

\newcommand{\aav}[1]{\left\langle #1 \right\rangle}
\newcommand{\sqrtsNN}{\sqrt{s_{\rm NN}}}

\renewcommand{\d}{{\rm d}}

\begin{document}

\title{Heavy ions and parton saturation from RHIC to LHC}
\author{A.~Dainese}
\institute{INFN - Laboratori Nazionali di Legnaro, Legnaro (Padova), Italy}
\maketitle
\begin{abstract}
The phenomenology of gluon saturation at small parton momentum fraction,
Bjorken-$x$, 
in the proton and in the nucleus
is introduced. The experimentally-accessible kinematic domains
at the nucleus--nucleus colliders RHIC and LHC are discussed.
Finally, the saturation hints emerging from measurements at RHIC and the
perspectives for LHC are described.
\end{abstract}

\section{Introduction: small-$x$ gluons in the proton and in the nucleus}

In the collinear factorization approach 
of perturbative QCD, 
the parton distribution functions (PDFs) of the proton are
determined through global fits obtained using 
the DGLAP 
scale evolution equations~\cite{Dokshitzer:1977sg,Gribov:1972ri,Altarelli:1977zs}. 
The HERA ep deep inelastic scattering (DIS) data on the proton structure function $F_2(x,Q^2)$ as a function of the parton momentum fraction Bjorken-$x$ and
of the squared momentum transfer $Q^2$, and, especially, the $Q^2$ slope,
$\partial F_2(x,Q^2)/\partial \ln Q^2$, in the
small-$x$, $3\times 10^{-5}\, \lsim x \lsim \,5\times 10^{-3}$,  
and small-$Q^2$ region, $1.5 \, \lsim Q^2 \lsim 10$~GeV$^2$,
set rather stringent constraints on the small-$x$ gluon
distribution $xg(x,Q^2)$. In this kinematic region, the gluon distribution
exhibits a strong rise towards low $x$ and
the agreement of the global fits with the HERA
$F_2(x,Q^2)$ data is not as good as it is at larger values of $x$
and $Q^2$~\cite{Martin:2003sk}. In particular, the gluon density $xg$ tends to rise
faster than what suggested by the data. This is due to the fact that 
the kernels of the DGLAP equations only describe splitting of one parton into
two or more, so that the resulting evolution is linear in the PDFs. 
At low $Q^2$, the small-$x$ gluon density may increase
to the point where gluon fusion, ${\rm gg\to g}$, becomes significant. 
Within the DGLAP framework, this phenomenology can be accounted for in an 
effective way by including nonlinear corrections in the evolution equations, that is, 
negative terms of order $\mathcal{O}(g^2)$, $\mathcal{O}(g^3)$, etc... that tame the evolution 
towards small $x$.
The first nonlinear corrections, the GLRMQ terms, were derived in Ref.~\cite{Gribov:1981ac,Mueller:1985wy}. 
A more accurate description of the small-$x$ nonlinearities is achieved in the
framework of $k_{\rm t}$-factorization, in which the BK equation~\cite{B,K} 
is used
to evolve the PDFs as a function of $x$ for fixed transverse momentum squared, $k_{\rm t}^2$, of the gluon.
Both approaches to nonlinear gluon dynamics, in DGLAP and in BK, 
suggest that one can expected potentially-measurable effects in pp collisions at 
LHC energy, for example in heavy-flavour production~\cite{heralhc}.

In the case of proton--nucleus and nucleus--nucleus collisions, 
where nuclei with large mass number A 
are involved, the nonlinear effects are enhanced by the larger density of 
gluons per unit transverse area of the colliding nuclei.
The high density of gluons at 
small $x$ and small $Q^2$ induces a suppression of the observed hard scattering
yields 
with respect to expectations based on a scaling with the
 number of binary nucleon--nucleon collisions. 
This reduction affects the kinematic region dominated by small-$x$ gluons: 
low transverse momentum $\pt$ and forward rapidity $y$, since, at leading 
order, we have $x\sim \pt\exp(-y)/\sqrtsNN$.
The effect, indicated as nuclear shadowing, 
is usually accounted for in terms of a
modification of the parton distribution functions of the nucleon
in the nucleus, $f_i^{\rm A}(x,Q^2)$, with respect to those of
the free nucleon, $f_i^{\rm N}(x,Q^2)$:
\begin{equation}
  \label{eq:shad}
  R_i^{\rm A}(x,Q^2) = \frac{f_i^{\rm A}(x,Q^2)}{f_i^{\rm N}(x,Q^2)}
\end{equation}
where $i = {\rm q_v,\,q_{sea},\,g}$ for valence quarks, 
sea quarks, and gluons. We have shadowing, $R_{\rm g}^{\rm A}<1$, 
for $x\lsim 5\times 10^{-2}$. However, as we will discuss in the following, the strength of the 
reduction is constrained by existing experimental data only for $x\gsim 10^{-3}$.

The use of nuclear-modified parton distribution functions allows
high-density effects at small $x$ to be accounted for within the framework of 
perturbative QCD collinear factorization. However, factorization 
is expected to break down when the gluon phase-space becomes {\it saturated}.
In these conditions, in the collision with an incoming projectile parton, 
the partons in the target nuclear wave function at small $x$ would 
act coherently, not independently as assumed with factorization.
In the limit, they may form 
a Colour Glass Condensate (CGC)~\cite{CGC}: a system, that can be describe in analogy to a 
spin glass, where gluons (colour charges) have large occupation number, as in a condensate.
The relevant parameter in the CGC is 
the so-called saturation scale $Q_{\rm S}^2$, defined as the 
scale at which the transverse area of the nucleus is completely saturated and gluons start to overlap.
This happens when the number of gluons, $\sim{\rm A}\,xg(x,Q^2_{\rm S})$, multiplied by the typical gluon size, $\sim 1/Q^2_{\rm S}$, is equal to the transverse area, $\sim\pi R^2_{\rm A}$. Thus:
\begin{equation}
\label{eq:qs}
Q^2_{\rm S}\sim\frac{{\rm A}\,xg(x,Q^2_{\rm S})}{\pi R^2_{\rm A}}\sim\frac{{\rm A}\,xg(x,Q^2_{\rm S})}{\rm A^{2/3}}\sim {\rm A}^{1/3}x^{-\lambda}\sim{\rm A}^{1/3} \big(\sqrtsNN\big)^\lambda e^{\lambda y}\,,~~~~{\rm with~}\lambda\approx0.3. 
\end{equation}
$Q^2_{\rm S}$ grows at forward rapidity, at high c.m.s. energy , 
and it is enhanced by a factor about 6 ($200^{1/3}$) in the Au or Pb nucleus, 
with respect to the proton.
Saturation affects the processes in the region $Q^2\lsim Q^2_{\rm S}$,
where gluon recombination dominates and factorization may start to become
invalid.

\section{Exploring the saturation region}

\begin{figure}[!t]
\begin{center}
\includegraphics[width=0.6\textwidth]{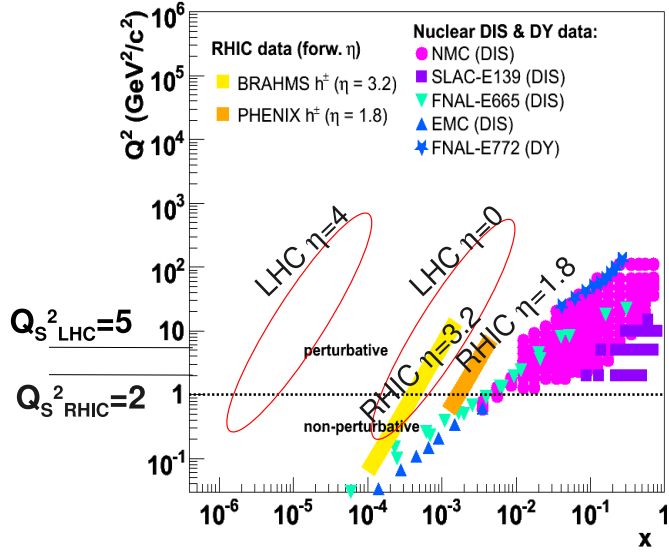}
\caption{The kinematic regions in $x$ and $Q^2$ explored by nuclear DIS 
and Drell-Yan experiments, by RHIC experiments, and by experiments in 
 preparation at LHC. Elaborated from a compilation in Ref.~\protect\cite{david}.}
\label{fig:accxQ2}
\end{center}
\end{figure}

Figure~\ref{fig:accxQ2}, elaborated from Ref.~\cite{david}, 
shows the experimental acceptances in the plane $(x,Q^2)$ for:
the nuclear DIS (lepton--nucleus) experiments NMC, SLAC-E139, FNAL-E665, EMC; 
the nuclear Drell-Yan (lepton--nucleus) experiment FNAL-E772; the RHIC 
(dAu) experiments BRAHMS and PHENIX; the experiments in preparation 
at LHC, ALICE, ATLAS, CMS, LHCb. 

The current knowledge of the nuclear modification of the PDFs is based on the
nuclear DIS data, reaching down to $x\gsim 10^{-3}$. As it can be seen from the 
figure, the LHC will give access to an unexplored small-$x$ domain of QCD.
There are several model extrapolations of the amount of nuclear shadowing 
in this region, with $R_g^{\rm Pb}(x,Q^2)$ ranging from 0.1 to 0.8 
at $x\sim 10^{-4}$ and $Q^2\sim 2~\gev^2$ (see e.g. Ref.~\cite{Eskola:2008ca}).
 
The estimated values of the saturation scale in heavy-ion collisions at 
RHIC and LHC are reported in the figure.
For a Au nucleus probed at RHIC energy, $\sqrtsNN=200~\gev$, 
the estimated saturation scale is $Q_{\rm S}^2\sim 2~\gev^2$: processes that 
involve gluons at $x<10^{-3}$--$10^{-2}$ are affected.
For a Pb nucleus probed at LHC energy, $\sqrtsNN=5.5~\tev$, 
the estimated saturation scale is $Q_{\rm S}^2\sim 5~\gev^2$: processes that 
involve gluons at $x<10^{-4}$--$10^{-3}$ are affected.
The line at $Q^2=1~\gev^2$ shows the lower limit of applicability of the 
perturbative QCD approach. At variance from RHIC, where the perturbative
region and the saturation region have little overlap, at the LHC it will 
be possible to explore the saturation region with perturbative probes, like
heavy quarks, and $\ccbar$ in particular. This means that discrepancies 
between charm production measurements close to the threshold
and perturbative predictions could signal the onset of saturation effects.
We will further discuss this point in Section~\ref{sec:lhchvq}.
Another very promising approach to the investigation of small-$x$ effects is
by measuring hard process (jets, heavy quarks, weak-interaction vector bosons)
in the forward rapidity region (see Section~\ref{sec:lhcforward}).

\section{Hints of saturation at RHIC}

\begin{figure}[htb]
\begin{center}
\includegraphics[width=0.49\textwidth]{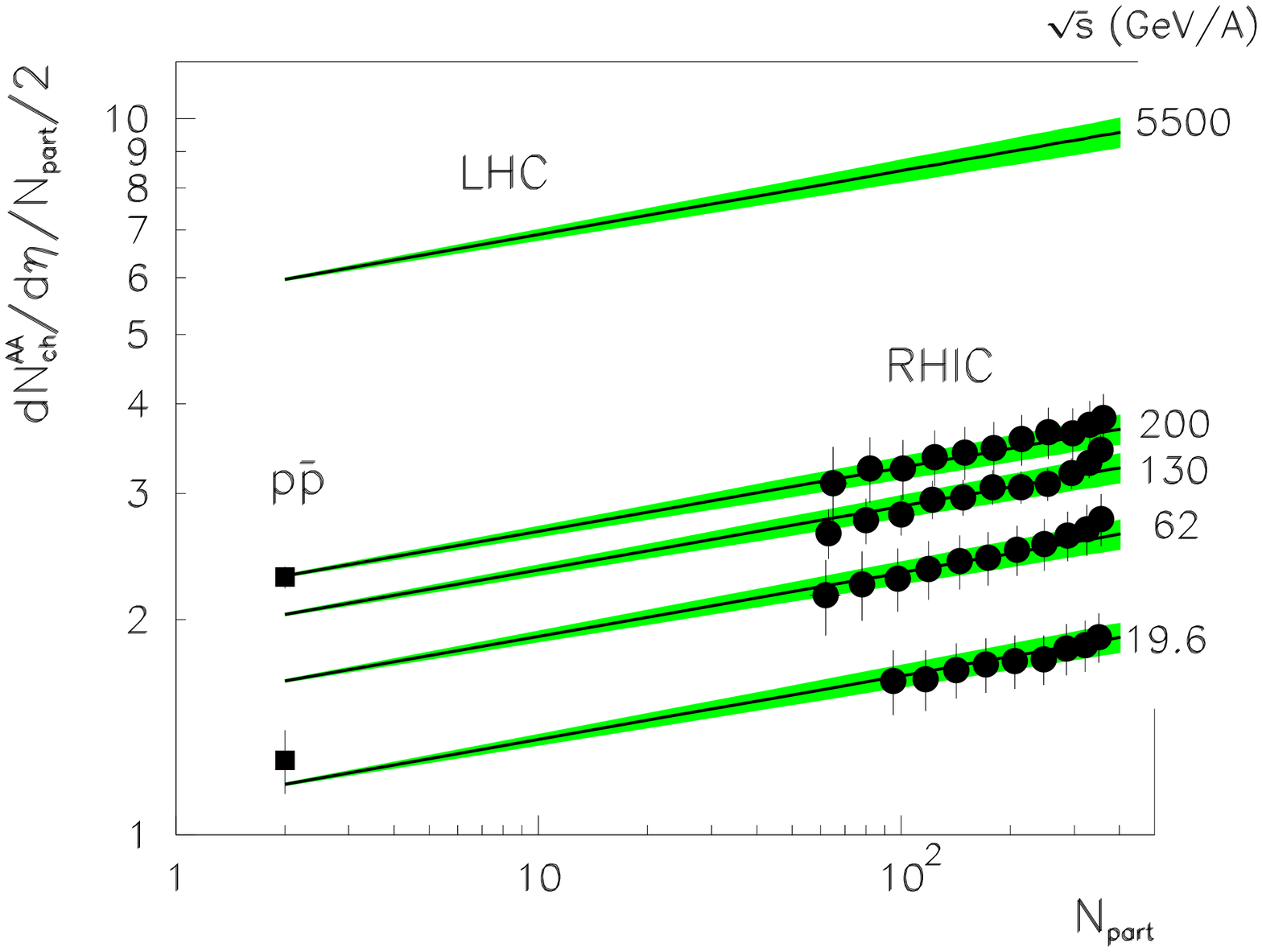}
\includegraphics[width=0.49\textwidth]{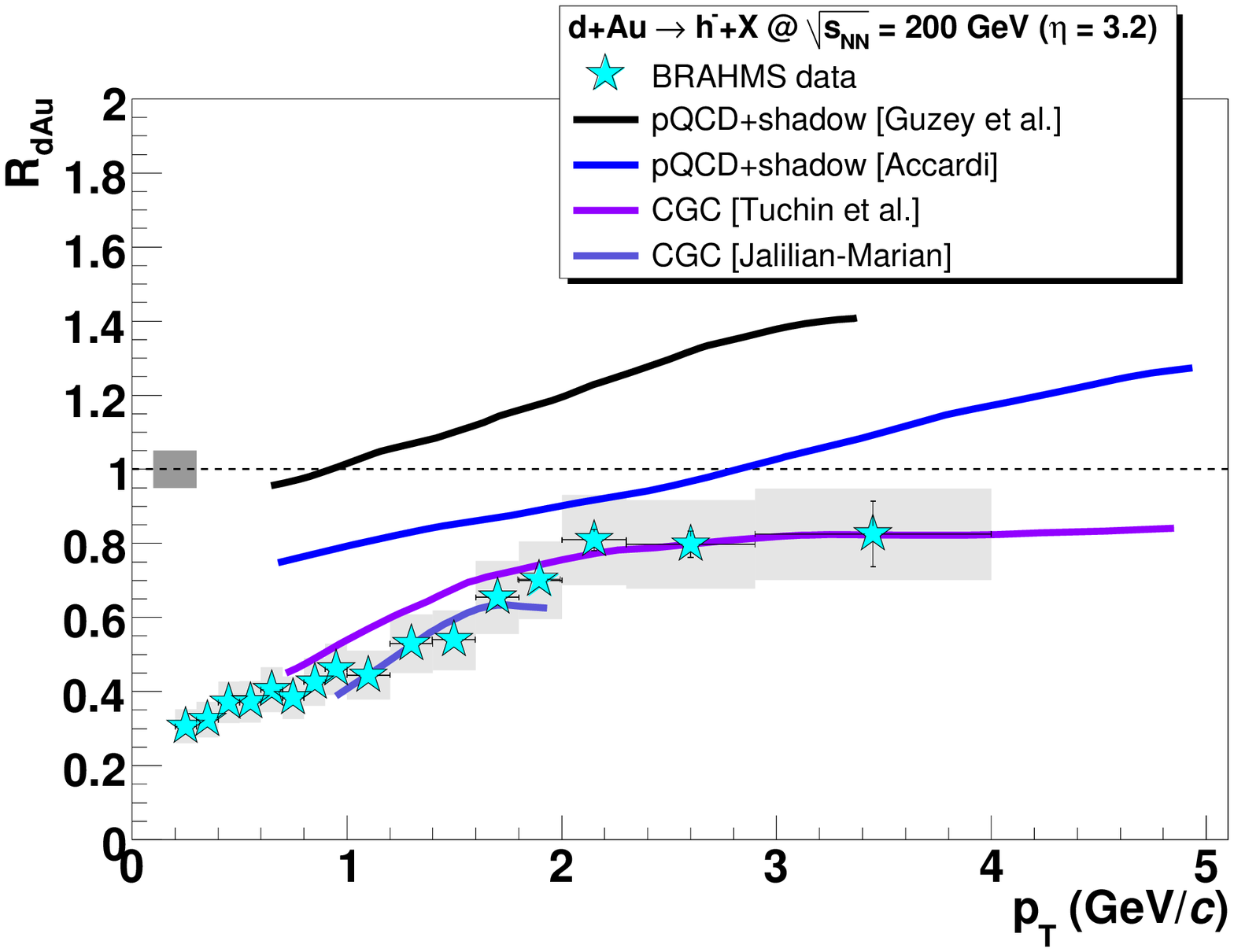}
\caption{Hints of saturation at RHIC. Left: Normalized $\dNdeta$ as a function of c.m.s. energy 
and centrality (given in terms of the number of nucleons participating in the collision, $N_{\rm part}$)  
measured by PHOBOS in Au--Au~\protect\cite{Back:2004je} compared with saturation predictions~\protect\cite{Armesto:2004ud}.
Right: Nuclear modification factor $R_{\rm dAu}(\pt)$ for negative hadrons at $\eta$~=~3.2 in dAu at $\sqrtsNN$ = 200 GeV: 
BRAHMS data~\protect\cite{Arsene:2004fa} compared to pQCD~\protect\cite{Guzey:2004zp,Accardi:2004fi} and 
CGC~\protect\cite{tuchin04,jamal04} predictions.}
\label{fig:rhic_sat}
\end{center}
\end{figure}

Two experimental observations in heavy-ion collisions at RHIC 
support the saturation
predictions of a reduced parton flux
in the incoming ions due to nonlinear QCD effects. 
On one hand, the measured 
hadron multiplicities (see e.g. Ref.~\cite{Back:2004je}),
$\dNdeta\approx 700$, are significantly lower than the $\dNdeta\approx 1000$ 
values predicted by 
minijet~\cite{Gyulassy:1994ew} or Regge~\cite{Capella:1992yb} models, but are well reproduced 
by CGC approaches~\cite{McLerran:1994vd}. 
 Assuming parton--hadron
duality, hadron multiplicities at mid-rapidity rise proportionally to 
$Q_{\rm S}^2$ times the transverse 
(overlap) area~\cite{Armesto:2004ud}, a feature that accounts naturally for the 
experimentally-observed factorization of 
$\sqrtsNN$- and centrality-dependences in $\dNdeta$ 
(Fig.~\ref{fig:rhic_sat}, left).
The second possible manifestation of CGC-like effects in the RHIC data is the BRAHMS observation~\cite{Arsene:2004fa} 
of suppressed yields of semi-hard hadrons ($\pt\approx 2$--4~GeV/$c$) in dAu relative to pp 
collisions at forward rapidities (up to $\eta\approx 3.2$, 
Fig.~\ref{fig:rhic_sat}, right).
Hadron production at such small angles is sensitive to partons in the Au nucleus 
with  $x^{\rm min} \sim \pt\exp(-\eta)/\sqrtsNN\sim 10^{-3}$~\cite{Guzey:2004zp}.
The observed nuclear modification factor, $R_{\rm dAu}\approx$ 0.8, cannot be reproduced by pQCD
calculations~\cite{Guzey:2004zp,Accardi:2004fi} that include the same nuclear shadowing
that describes the dAu data at $\eta=0$, but can be described by CGC approaches that parametrise 
the Au nucleus as a saturated gluon wavefunction~\cite{tuchin04,jamal04}.

\section{Perspectives for LHC}
\label{sec:lhc}

\subsection{Accessing the small-$x$ region with hard processes at forward rapidity}
\label{sec:lhcforward}

The four LHC experiments ---i.e. the two general-purpose and high-luminosity ATLAS and CMS 
detector systems as well as the heavy-ion-dedicated ALICE and the heavy-flavour-oriented LHCb 
experiments--- have all detection capabilities in the forward direction very well adapted for the study 
of low-$x$ QCD phenomena with hard processes in collisions with proton and ion beams (see e.g. Ref.~\cite{davidlhc} for more details):
\begin{itemize}
\item Both CMS and ATLAS feature hadronic calorimeters in the range 
3$<|\eta|<$5 which allow them to measure jet cross-sections at very forward rapidities.
Both experiments feature also zero-degree calorimeters (ZDC, $|\eta|\gsim 8.5$ for neutrals), 
which are a basic tool for neutron-tagging ``ultra-peripheral'' Pb--Pb photoproduction interactions. 
CMS has an additional electromagnetic/hadronic calorimeter (CASTOR, $5.3<|\eta|<6.7$)
and shares the interaction point with the TOTEM experiment providing two extra 
trackers at very forward rapidities (T1, $3.1 <|\eta|< 4.7$, and T2, $5.5<|\eta|<6.6$) 
well-suited for DY measurements.
\item The ALICE forward muon spectrometer at $2.5<\eta < 4$, can reconstruct
$J/\psi$ and $\Upsilon$ (as well as $Z^0$) in the di-muon channel, as well as statistically measure 
single inclusive heavy-quark production via semi-leptonic (muon) decays. ALICE counts 
also on ZDCs in both sides of the interaction point for forward neutron triggering 
of Pb--Pb photoproduction processes.
\item LHCb is a single-arm spectrometer covering rapidities $1.8<\eta<4.9$,  with 
very good particle identification capabilities designed to accurately reconstruct charm and beauty 
hadrons. The detector is also well-suited to measure jets, $Q\overline{Q}$ and $Z^0\to\mu\mu$ production in the forward hemisphere.
\end{itemize}

\subsection{Probing small-$x$ gluons with heavy quarks}
\label{sec:lhchvq}

\begin{figure}[!t]
\begin{center}
\includegraphics[width=0.34\textwidth]{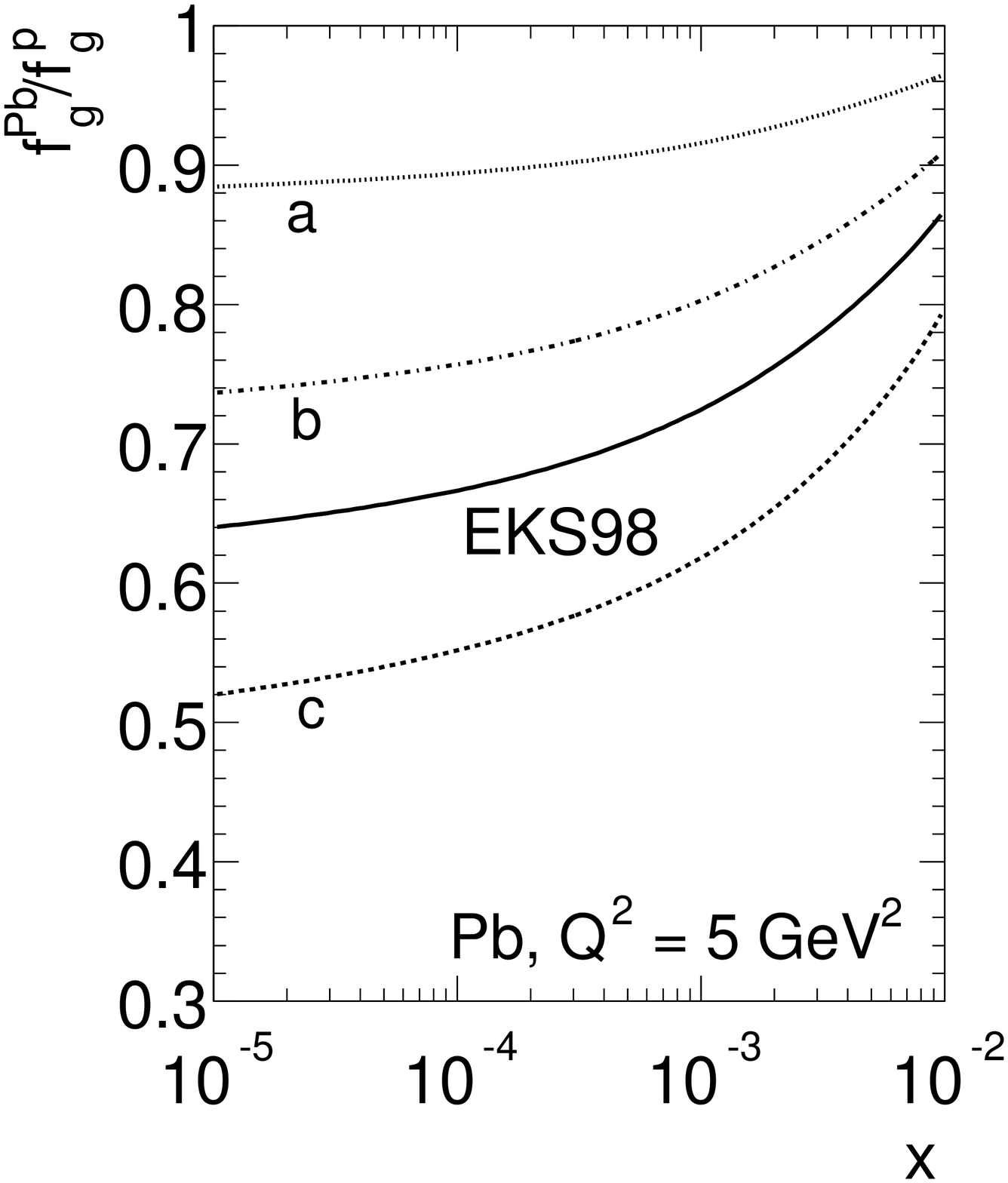}
\includegraphics[width=0.44\textwidth]{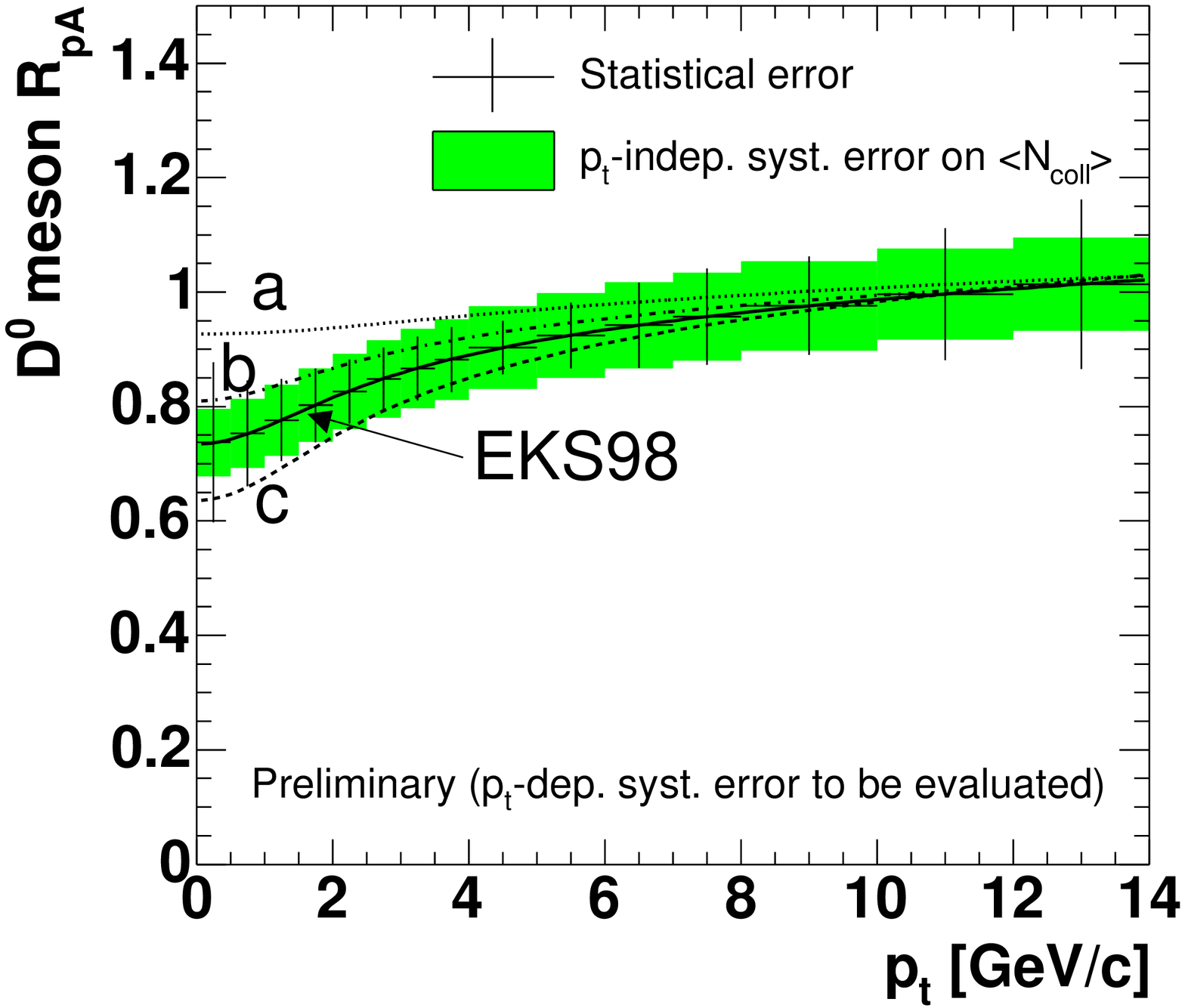}
\caption{Charm production in p--Pb at LHC. 
         Left: EKS98~\protect\cite{Eskola:1998df} parametrisation  for the 
         modification of the gluon PDF 
         in a Pb nucleus at $Q^2=5~\gev^2\simeq 4\,m_{\rm c}^2$,
         along with three other different scenarios. 
         Right: corresponding
         $R_{\rm pA}^{\rm D}$ in p--Pb at $\sqrtsNN=8.8~\tev$ and expected 
         sensitivity of the ALICE experiment with the $\rm D^0\to K^-\pi^+$ 
        measurement at central rapidity ($|\eta|<0.9$), 
         with one year of data taking at nominal LHC luminosity~\protect\cite{grosso}.}
\label{fig:smallx}
\end{center}
\end{figure}

As already mentioned, at LHC it will be possible to probe the saturation 
region with perturbative probes, such as heavy quarks.
The $x$ regime 
relevant for charm production in heavy-ion collisions at LHC 
($x\gsim 2m_{\rm c}\exp(-y)/\sqrtsNN$) extends down to $x\sim 10^{-4}$ 
already at central rapidity $y=0$ and down to $x\sim 10^{-6}$ at forward
rapidity $y\approx 4$~\cite{Carrer:2003ac}.
Charm (and beauty) production cross sections at small $\pt$ and forward
rapidity are thus 
expected to be significantly affected by parton dynamics in the 
small-$x$ region. 
As an example,
the EKS98 parametrisation~\cite{Eskola:1998df} of the PDFs nuclear 
modification, 
shown in Fig.~\ref{fig:smallx} (centre) for $Q^2=5~\gev^2$,
predicts a  reduction of the charm (beauty) cross section at NLO
of about 35\% (20\%) in \mbox{Pb--Pb} at 5.5~TeV and 15\% (10\%) in \mbox{pPb}
at 8.8~TeV~\cite{Carrer:2003ac}.

The comparison of heavy-quark production in 
pp and pPb collisions (where final-state effects, such as parton 
energy loss, are not expected to be 
present) 
is regarded as a sensitive tool to probe nuclear PDFs at LHC energy.
The ratio of invariant-mass spectra of dileptons from heavy-quark decays 
in pPb and pp collisions would measure 
the nuclear modification $R_{\rm g}^{\rm Pb}$~\cite{Accardi:2004be}. 
Another promising observable in this respect is the nuclear modification 
factor of the D meson $\pt$ distribution, defined as:
\begin{equation}
R^{\rm D}_{\rm pA(AA)}(\pt,\eta)=
{1\over \aav{N_{\rm coll}}} \times 
{\d^2 N^{\rm D}_{\rm pA(AA)}/\d\pt\d\eta \over 
\d^2 N^{\rm D}_{\rm pp}/\d\pt\d\eta}\,.
\label{eq:RAA}
\end{equation}
In Fig.~\ref{fig:smallx} (right) we show the sensitivity of 
$R^{\rm D}_{\rm pPb}$ to 
different shadowing scenarios, obtained by varying the modification of the 
PDFs in the Pb nucleus 
(displayed, 
for gluons, by the curves labeled `a', `b', `c' and `EKS98'
in the left panel of the same figure). The ALICE experiment will 
be able to measure D meson production down to almost zero transverse momentum,
at central rapidity~\cite{Alessandro:2006yt}. As shown by the projected 
experimental uncertainties
on the $\rm D^0$ nuclear modification factor in pPb,
reported in the left panel of Fig.~\ref{fig:smallx}, this measurement 
is expected to be sensitive to the level of nuclear shadowing
at LHC. 

\begin{figure}[!t]
\begin{center}
\includegraphics[width=0.43\textwidth]{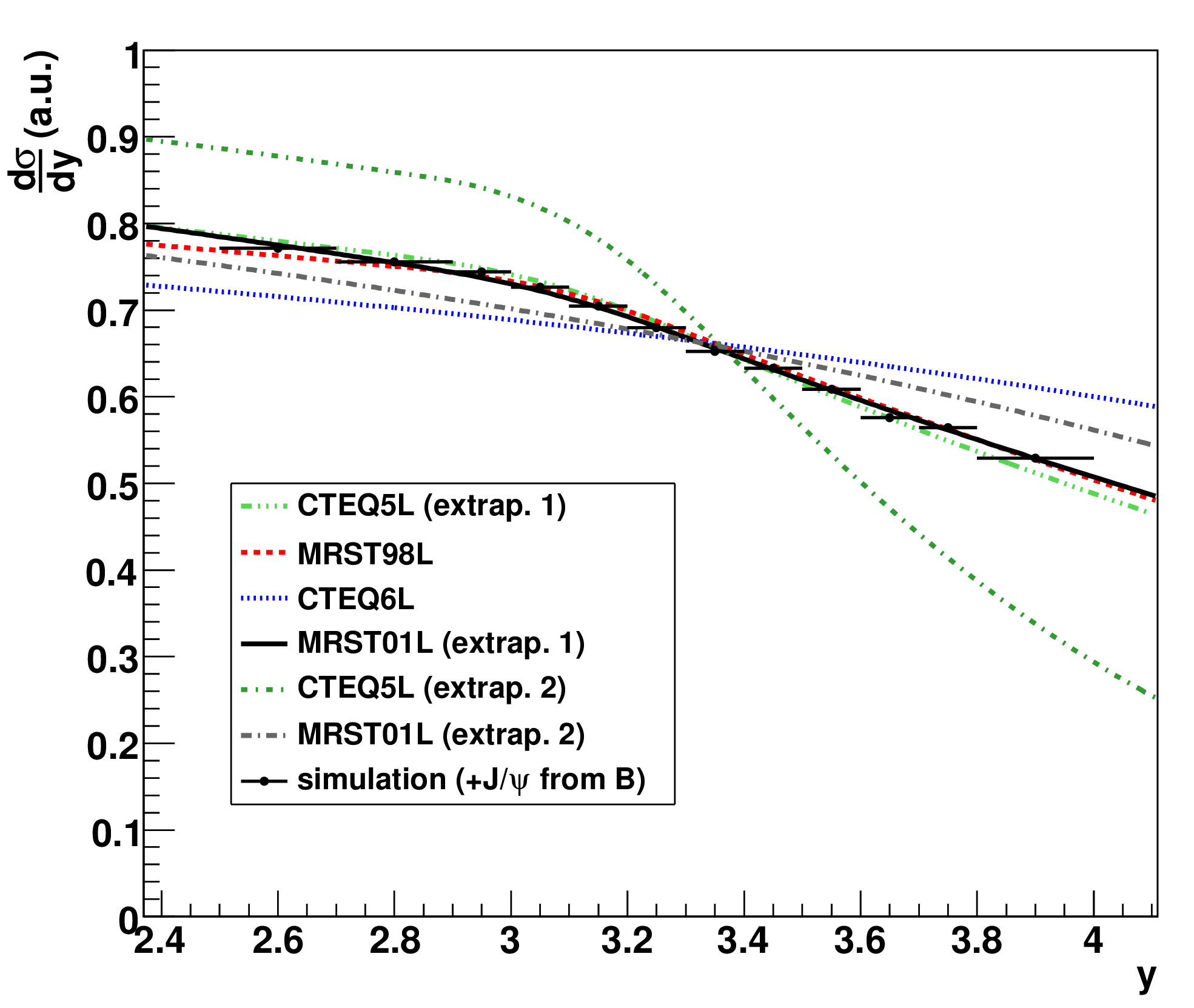}
\caption{$J/\psi$ production in pp at LHC. 
         Cross section as a function of rapidity as predicted using
         different PDF sets (details in the text) , compared to the 
         projected precision of the measurement of the ALICE experiment~\protect\cite{stocco}.}.
\label{fig:stocco}
\end{center}
\end{figure}

Charmonium production at low $\pt$ and forward rapidity 
is another promising probe of small-$x$ gluons at LHC.
All four LHC experiments are expected to have good capability for 
$J/\psi$ reconstruction in the central and in the forward rapidity region.
In particular, ALICE will provide a 
measurement via di-muons in $2.5<y<4$ down to $\pt\approx 0$~\cite{Alessandro:2006yt}, which probes
the poorly-known region $x<10^{-5}$ where current PDF parametrisations 
have large uncertainties. Figure~\ref{fig:stocco} shows the $J/\psi$ 
rapity-differential cross section at NLO from the Color Evaporation Model~\cite{hphvq}, in $2.5<y<4$ with different PDF sets, compared
to the projected precision of the ALICE measurement~\cite{stocco}.


\section{Summary}
We have discussed how gluons nonlinear evolution and the phenomenology of
saturation are expected to set on at small $x$ in the hadrons, and how these
effects are enhanced by the higher transverse gluon density in large nuclei.
The study of this
almost unexplored regime can provide fundamental insight on the 
high-energy limit of QCD. 
We have described the experimental indications 
of the onset of saturation in heavy-ion collisions at RHIC.
The LHC, as a heavy-ion collider, will be a unique laboratory for the 
investigation of the saturation regime with perturbative probes, 
such as forward rapidity hard processes and heavy quarks at 
low momentum and/or forward rapidity.

\paragraph{Acknowledgements:} I am grateful to the organizers of the 
International Symposium on Multi-particle Dynamics 2008 for the invitation and
support. I warmly thank D.~d'Enterria for stimulating discussions and for 
providing useful material for the preparation of this contribution, and K.~Kutak
for valuable comments on the manuscript.

\begin{footnotesize}
\bibliographystyle{ismd08} 
{\raggedright
\bibliography{ismd08}

\providecommand{\etal}{et al.\xspace}
\providecommand{\href}[2]{#2}
\providecommand{\coll}{Coll.}
\catcode`\@=11
\def\@bibitem#1{%
\ifmc@bstsupport
  \mc@iftail{#1}%
    {;\newline\ignorespaces}%
    {\ifmc@first\else.\fi\orig@bibitem{#1}}
  \mc@firstfalse
\else
  \mc@iftail{#1}%
    {\ignorespaces}%
    {\orig@bibitem{#1}}%
\fi}%
\catcode`\@=12
\begin{mcbibliography}{10}

\bibitem{Dokshitzer:1977sg}
Y.~L. Dokshitzer,
\newblock Sov. Phys. JETP{} {\bf 46},~641~(1977)\relax
\relax
\bibitem{Gribov:1972ri}
V.~N. Gribov and L.~N. Lipatov,
\newblock Sov. J. Nucl. Phys.{} {\bf 15},~438~(1972)\relax
\relax
\bibitem{Altarelli:1977zs}
G.~Altarelli and G.~Parisi,
\newblock Nucl. Phys.{} {\bf B126},~298~(1977)\relax
\relax
\bibitem{Martin:2003sk}
A.~D. Martin, R.~G. Roberts, W.~J. Stirling, and R.~S. Thorne,
\newblock Eur. Phys. J.{} {\bf C35},~325~(2004)\relax
\relax
\bibitem{Gribov:1981ac}
L.~V. Gribov, E.~M. Levin, and M.~G. Ryskin,
\newblock Nucl. Phys.{} {\bf B188},~555~(1981)\relax
\relax
\bibitem{Mueller:1985wy}
A.~H. Mueller and J.-w. Qiu,
\newblock Nucl. Phys.{} {\bf B268},~427~(1986)\relax
\relax
\bibitem{B}
I.~Balitsky,
\newblock Nucl. Phys.{} {\bf B463},~99~(1996)\relax
\relax
\bibitem{K}
Y.~Kovchegov,
\newblock Phys. Rev.{} {\bf D60},~034008~(1999)\relax
\relax
\bibitem{heralhc}
J.~Baines {\em et al.}~(2006).
\newblock \href{http://www.arXiv.org/abs/hep-ph/0601164}{{\tt
  hep-ph/0601164}}\relax
\relax
\bibitem{CGC}
E.~Iancu and R.~Venugopalan,
\newblock World Scientific, Singapore{}.
\newblock \href{http://www.arXiv.org/abs/hep-ph/0303204}{{\tt
  hep-ph/0303204}}\relax
\relax
\bibitem{david}
D.~G. d'Enterria,
\newblock Eur. Phys. J.{} {\bf A31},~816~(2007).
\newblock \href{http://www.arXiv.org/abs/hep-ex/0610061}{{\tt
  hep-ex/0610061}}\relax
\relax
\bibitem{Eskola:2008ca}
K.~J. Eskola, H.~Paukkunen, and C.~A. Salgado,
\newblock JHEP{} {\bf 07},~102~(2008).
\newblock \href{http://www.arXiv.org/abs/0802.0139}{{\tt 0802.0139}}\relax
\relax
\bibitem{Back:2004je}
B.~B. Back {\em et al.},
\newblock Nucl. Phys.{} {\bf A757},~28~(2005).
\newblock \href{http://www.arXiv.org/abs/nucl-ex/0410022}{{\tt
  nucl-ex/0410022}}\relax
\relax
\bibitem{Gyulassy:1994ew}
M.~Gyulassy and X.-N. Wang,
\newblock Comput. Phys. Commun.{} {\bf 83},~307~(1994).
\newblock \href{http://www.arXiv.org/abs/nucl-th/9502021}{{\tt
  nucl-th/9502021}}\relax
\relax
\bibitem{Capella:1992yb}
A.~Capella, U.~Sukhatme, C.-I. Tan, and J.~Tran Thanh~Van,
\newblock Phys. Rept.{} {\bf 236},~225~(1994)\relax
\relax
\bibitem{McLerran:1994vd}
L.~D. McLerran and R.~Venugopalan,
\newblock Phys. Rev.{} {\bf D50},~2225~(1994).
\newblock \href{http://www.arXiv.org/abs/hep-ph/9402335}{{\tt
  hep-ph/9402335}}\relax
\relax
\bibitem{Armesto:2004ud}
N.~Armesto, C.~A. Salgado, and U.~A. Wiedemann,
\newblock Phys. Rev. Lett.{} {\bf 94},~022002~(2005).
\newblock \href{http://www.arXiv.org/abs/hep-ph/0407018}{{\tt
  hep-ph/0407018}}\relax
\relax
\bibitem{Arsene:2004fa}
I.~Arsene {\em et al.},
\newblock Nucl. Phys.{} {\bf A757},~1~(2005).
\newblock \href{http://www.arXiv.org/abs/nucl-ex/0410020}{{\tt
  nucl-ex/0410020}}\relax
\relax
\bibitem{Guzey:2004zp}
V.~Guzey, M.~Strikman, and W.~Vogelsang,
\newblock Phys. Lett.{} {\bf B603},~173~(2004).
\newblock \href{http://www.arXiv.org/abs/hep-ph/0407201}{{\tt
  hep-ph/0407201}}\relax
\relax
\bibitem{Accardi:2004fi}
A.~Accardi,
\newblock Acta Phys. Hung.{} {\bf A22},~289~(2005).
\newblock \href{http://www.arXiv.org/abs/nucl-th/0405046}{{\tt
  nucl-th/0405046}}\relax
\relax
\bibitem{tuchin04}
D.~Kharzeev, Y.~Kovchegov, and K.~Tuchin,
\newblock Phys. Lett.{} {\bf B599},~23~(2004)\relax
\relax
\bibitem{jamal04}
J.~Jalilian-Marian,
\newblock Nucl. Phys.{} {\bf A748},~664~(2005).
\newblock \href{http://www.arXiv.org/abs/nucl-th/0402080}{{\tt
  nucl-th/0402080}}\relax
\relax
\bibitem{davidlhc}
D.~G. d'Enterria~(2008).
\newblock \href{http://www.arXiv.org/abs/arXiv:0806.0883}{{\tt
  arXiv:0806.0883}}\relax
\relax
\bibitem{Carrer:2003ac}
N.~Carrer and A.~Dainese~(2003).
\newblock \href{http://www.arXiv.org/abs/hep-ph/0311225}{{\tt
  hep-ph/0311225}}\relax
\relax
\bibitem{Eskola:1998df}
K.~J. Eskola, V.~J. Kolhinen, and C.~A. Salgado,
\newblock Eur. Phys. J.{} {\bf C9},~61~(1999).
\newblock \href{http://www.arXiv.org/abs/hep-ph/9807297}{{\tt
  hep-ph/9807297}}\relax
\relax
\bibitem{Accardi:2004be}
A.~Accardi {\em et al.}~(2004).
\newblock \href{http://www.arXiv.org/abs/hep-ph/0308248}{{\tt
  hep-ph/0308248}}\relax
\relax
\bibitem{Alessandro:2006yt}
B.~Alessandro {\em et al.},
\newblock J. Phys.{} {\bf G32},~1295~(2006)\relax
\relax
\bibitem{grosso}
R.~Grosso,
\newblock PhD Thesis, Universit\`a di Trieste{}~(2004)\relax
\relax
\bibitem{stocco}
D.~Stocco,
\newblock PhD Thesis, Universit\`a di Torino{}~(2008)\relax
\relax
\bibitem{hphvq}
M.~Bedjidian {\em et al.}~(2003).
\newblock \href{http://www.arXiv.org/abs/hep-ph/0311048}{{\tt
  hep-ph/0311048}}\relax
\relax
\end{mcbibliography}


\providecommand{\etal}{et al.\xspace}
\providecommand{\href}[2]{#2}
\providecommand{\coll}{Coll.}
\catcode`\@=11
\def\@bibitem#1{%
\ifmc@bstsupport
  \mc@iftail{#1}%
    {;\newline\ignorespaces}%
    {\ifmc@first\else.\fi\orig@bibitem{#1}}
  \mc@firstfalse
\else
  \mc@iftail{#1}%
    {\ignorespaces}%
    {\orig@bibitem{#1}}%
\fi}%
\catcode`\@=12
\begin{mcbibliography}{1}

\bibitem{feyn}
R.~Feynman,
\newblock {\em Photon--Hadron Interactions}.
\newblock Benjamin, New York, 1972\relax
\relax
\bibitem{art1}
{ UA5} Collaboration, G.~J. Alner {\em et al.},
\newblock Z. Phys.{} {\bf C33},~1~(1986)\relax
\relax
\bibitem{art2}
{ CDF} Collaboration, F.~Abe {\em et al.},
\newblock Phys. Rev.{} {\bf D41},~2330~(1990)\relax
\relax
\bibitem{art3}
B.~Andersson, G.~Gustafson, and J.~Samuelsson,
\newblock Nucl. Phys.{} {\bf B467},~443~(1996)\relax
\relax
\end{mcbibliography}
}
\end{footnotesize}
\end{document}